\documentclass{ws-p8-50x6-00}

\begin{document}

\title{Multiplicity Fluctuations and Bose-Einstein Correlations in DIS at HERA} 

\author{M. Adamus$^\dagger$, S. Chekanov$^\ddagger$, K. Olkiewicz$^\ast$ and L. Zawiejski$^\ast$}

\address{$^\dagger$ Institute for Nuclear Studies, Warsaw, Poland \\
$\ddagger$ Argonne National Laboratory, Argonne, IL, USA \\
$\ast$ Institute of Nuclear Physics, Cracow, Poland} 


\maketitle

\vspace*{-0.1cm}
{\hspace*{1.cm}  presented by L. Zawiejski for ZEUS Collaboration}
\vspace*{0.2cm}
\abstracts{
Results of the recent studies of the multiplicity fluctuations and 
Bose-Einstein correlations (BEC) in deep-inelastic scattering (DIS) at large Q$^2$
are reviewed. The measurements were done with the ZEUS detetor at HERA.}
\vspace*{-0.9cm}
\section{Introduction}
Various aspects of multihadron production can be revealed through multiplicity
fluctuations and particle correlations studies. The fluctuations are sensitive to 
soft particle dynamics and their measurements allow to test perturbative QCD \cite{och}
and local parton hadron duality (LPHD) hypothesis \cite{ph1}. 
In BEC studies \cite{berep2}, the correlation function gives information about 
the shape, size and lifetime of the boson emitter source. \\ 
In the fluctuations studies, the method of normalized
factorial moments, $F_{q}$, was used \cite{fm1}.
The moments of order $q$ were calculated by counting
$n$, the number of charged particles in a restricted region    
of phase space, $\Omega$:
\vspace*{-0.1cm}
\begin{equation}
F_q(\Omega) \; = \; <n>^{-q}\; <n(n-1)\ldots (n-q+1)>,
\end{equation}
where $< \ldots >$ denotes averaging over the sample and $\Omega$
is defined
either as a polar-angle ring around the jet axis
or as an upper limit on the particle transverse or absolute momentum calculated 
w.r.t. the jet axis. 
\vspace*{-0.2cm}
\section{Angular multiplicity fluctuations}
The factorial
moments were measured as a function of the ring width $\theta$ 
in the corresponding cone with half opening angle $\Theta_{0}$ \cite{zeu}.
\begin{figure}[h]
\begin{center}
\epsfxsize=29pc 
\epsfysize=19pc
\epsfbox{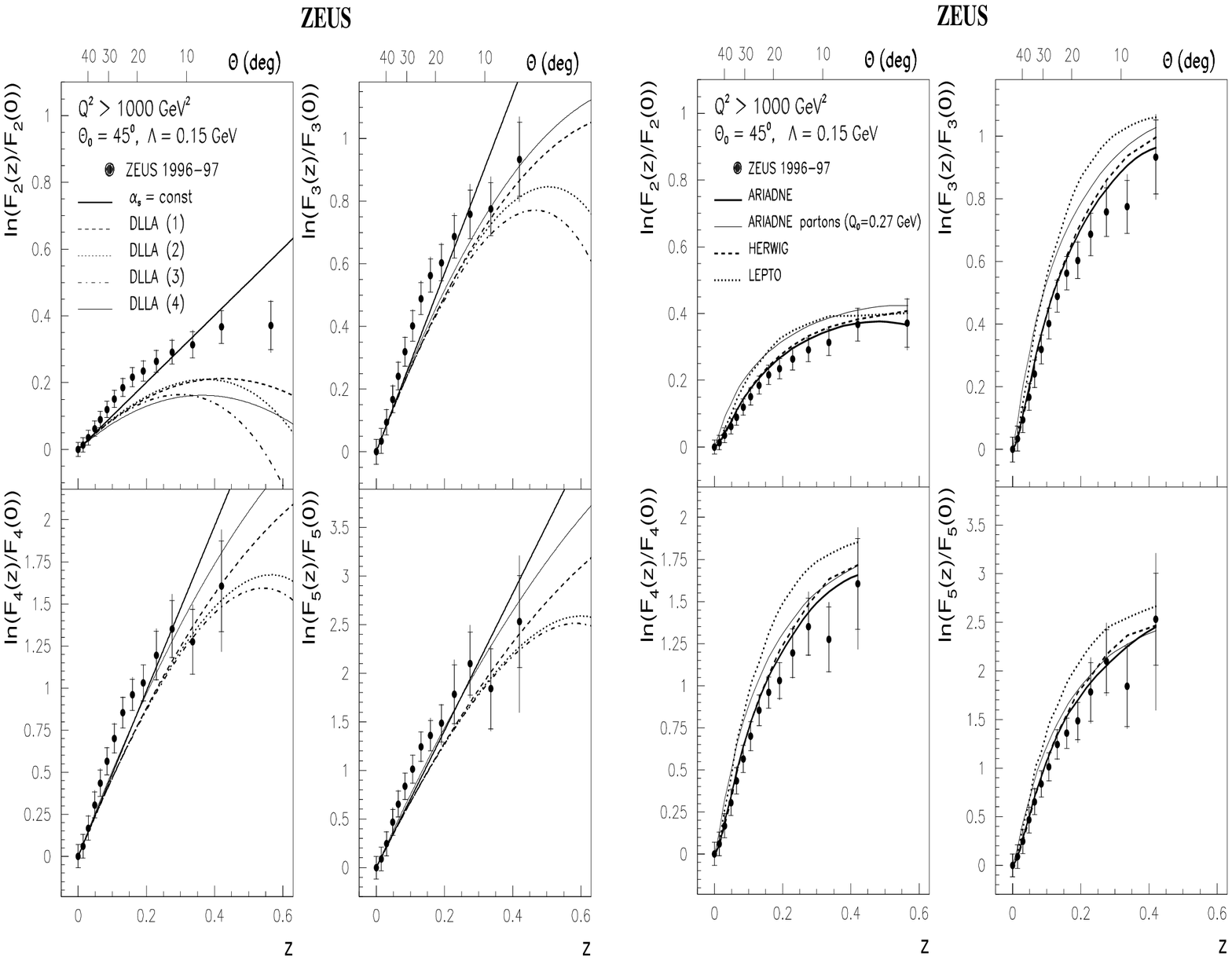} 
\vspace*{-0.9cm}
\caption{(Left): Comparison of the factorial moments with QCD+LPHD calculations.
The curve DLLA (4) includes
a correction from the MLLA in the calculation of $D_{q}$;
(Right): Comparison of the factorial moments
with different MC models at hadron and parton level.}
\vspace*{-0.25cm}
\end{center}
\end{figure}
Substituting $\theta$ with a scaling variable, 
 $z=\ln(\Theta_{0}/ \theta) / \ln(E \Theta_{0}/ \Lambda)$, the 
experimental results can be compared with analytic calculations where 
$E$ is energy of an outgoing quark radiating the gluons. 
The QCD + LPHD prediction for $F_{q}$ is:
\begin{equation}
\ln \frac{F_q(z)}{F_q(0)}=z\;(1-D_q)(q-1)\;\ln(E \Theta_{0}/ \Lambda),
\label{eq:mf2}
\end{equation}
where $D_{q}$ are the R$\acute{e}$nyi dimensions \cite{red}  
which can be calculated either
in a fixed or in a running-coupling
regime (see \cite{och}) of the Double Leading Log Approximation (DLLA).
For independent particle production:
$D_{q} = 1 $ and $F_{q}(z) = F_{q}(0)$.
Figure 1(left) compares the factorial moments for the DIS
data with the QCD predictions.
A significant disagreement with the data was found. 
Figure 1(right) shows comparison with
Monte Carlo predictions at hadron and parton levels.
All MC models reproduce the trends seen in the data. 
For consistency with the LPHD picture,
the parton cascade was cut-off at $Q_{0}$ = 0.27 GeV, which is close 
to $\Lambda = 0.22 $ GeV.
For higher order moments, the parton level is closer to the data
than the analytic calculations.
\vspace*{-0.2cm}
\section{Multiplicity fluctuations in limited momentum space}
According to QCD+LPHD predictions \cite{pt1}, the normalized factorial moments are expected
to behave as 
\begin{equation}
\hspace*{-1.cm} F_q(p_t^{cut})  \simeq 
1 + \frac{q(q-1)}{6}\>  \frac{\ln(p_t^{cut}/Q_0)}{\ln(E/Q_0)}, \hspace*{0.6cm}  F_q(p^{cut}) \simeq const(q) > 1,
\label{eq:mf3}
\end{equation} 
when particles are restricted in either the transverse momentum $p_{t} < p_{t}^{cut}$
or in spherical momentum $p < p^{cut}$.
If $p_{t}^{cut} \rightarrow Q_{0}$, then all $F_{q} \rightarrow 1 $  and the 
multiplicity distribution approaches Poissonian distribution due to the coherence effect
of soft gluons in a parton cascade. This is not the case for soft gluons with spherical 
momentum cut.
Figure 2 shows $p_{t}^{cut}$ and $p^{cut}$ moments \cite{zeu} together   
with Monte Carlo predictions 
at the hadron and parton levels. The parton level is consistent with 
analytic calculations.
For $p_{t} < 1$ GeV, Fig. 2(left), all moments rise 
in contradiction to the analytic QCD predictions.
The MC results for hadrons show similar trends to the data. The similar behaviour 
is found for $p^{cut}$ factorial moments
for small momentum cut off, Fig. 2(right). 
The observed disagreement between QCD+LPHD and the measurements is on qualitative level.      
\begin{figure}[t]
\begin{center}
\epsfxsize=26pc 
\epsfysize 22pc
\epsfbox{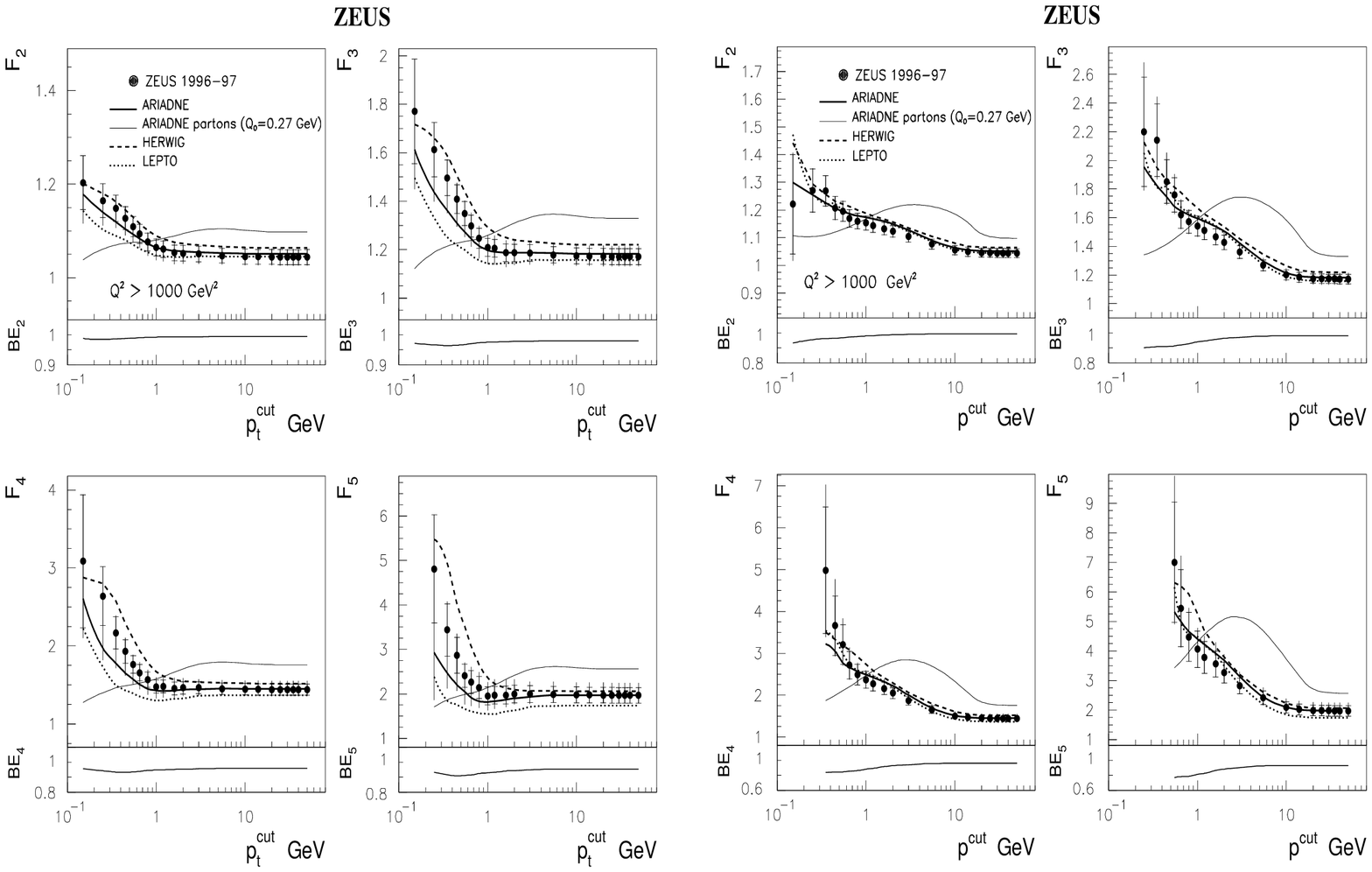} 
\vspace*{-2.2cm}
\caption{(Left): Comparison of the factorial moments calculated as a function of $p_{t}^{cut}$
with different MC models
at hadron level and parton-level ARIADNE, which represents analytic
calculations and is consistent with LPHD; (Right): The same for $p^{cut}$ factorial moments.}
\vspace*{-0.25cm}
\end{center}
\end{figure}
\vspace*{-0.2cm}
\section{Bose-Einstein correlations at large Q$^2$}
Bose-Einstein correlations between pairs of identical bosons can
be expressed by the two-particle correlation function of the Lorentz-invariant four-momentum
transfer $Q_{12} = \sqrt{-(p_1-p_2)^2}$ ($p_1$, $p_2$ are the four-momenta of the particles) : 
$R(Q_{12}) = \rho(Q_{12})/\rho_0(Q_{12})$, where $\rho_0(Q_{12})$ represents the 
two-particle density without the Bose-Einstein effect. 
\begin{figure}[hbt]
\begin{center}
\epsfxsize=26pc 
\epsfysize 10pc
\epsfbox{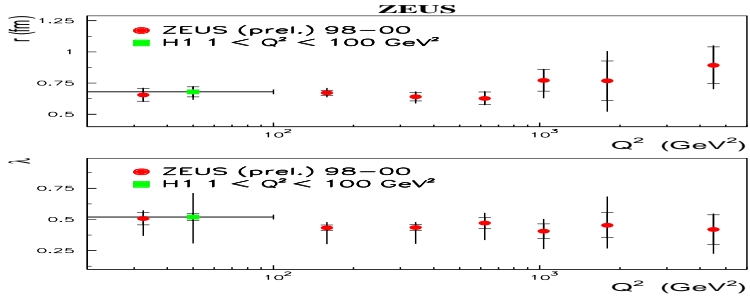} 
\vspace*{-0.3cm}
\caption{The radius $r$ and strength $\lambda$ of BEC as a function of Q$^2$.}
\vspace*{-0.8cm}
\end{center}
\end{figure}
\vspace*{0.2cm}
After the corrections for dynamical correlations and detector effects 
(Monte Carlo without BEC), the correlation function R was fitted to the 
following expression:
\vspace*{-0.15cm}
\begin{equation}
R(Q_{12}) = \kappa (1+\epsilon Q_{12})(1+\lambda e^{-r^2Q_{12}^2}), 
\end{equation}
where $r$ estimates the size of the two-boson emitter which is taken to be of Gaussian shape,
 $\lambda$ measures the BEC strength, factor $1+\epsilon Q_{12}$ takes into acount 
possible long-range momentum correlations and $\kappa$ is the normalization constant. \\
The BEC were studied to test of the energy 
dependence of r and $\lambda$. Figure 3 shows 
r and $\lambda$
as a function of four-momentum transfer squared Q$^2$.
No Q$^2$ dependence was found for r and $\lambda$.
The H1 DIS results \cite{hh1} at lower Q$^2$
are consistent with ZEUS data.
\vspace*{0.2cm}
\vspace*{-0.5cm}
\section*{Acknowledgments}
\vspace*{-0.2cm}
I would like to thank the DESY Directorate for financial support. \\
Thanks to 
J. Chwastowski for a
careful reading of the manuscript and useful remarks.

\end{document}